# Theory of off-normal incidence ion sputtering of surfaces of type $A_xB_{1-x}$ and a conformal map method for stochastic continuum models.

Oluwole Oyewande[1,*] and Boluwatife Adeoti[2]

[1] Department of Physics, University of Ibadan, Ibadan, Nigeria
[2] Department of Physics and Solar Energy, Bowen University, Iwo, Nigeria
*E-mail: oe.oyewande@mail.ui.edu.ng.

**Abstract**

Bradley et. al. recently provided an explanation of nanodot and defect-free ordered ripple production from binary compounds, for normal and oblique incidence ion sputtering respectively, by the inclusion of the effect of the preferential sputtering of one type of atom relative to the other type on the surfaces of binary compounds. In this paper, we propose an extended anisotropic model of such surfaces of type $A_xB_{1-x}$ that addresses anisotropy in the variations of the specie-composition as well. We show that this model gives the anisotropic Cuerno-Barabasi model as x approaches unity, and analyze the general properties of the model. Further, the complexity of the solutions of non-linear higher-order differential equations in general has led to the creation of great number of highly technical and computationally intensive numerical schemes. We introduce a simple conformal map method which allows for a fast and accurate simulation of the dynamical evolution of ion-sputtered surfaces based on any stochastic continuum model. An optimization algorithm for an efficient application of the scheme is also introduced. In this scheme the noise term has a physical meaning which allows one to go beyond the usual white noise approximation by actually being able to assign physical parameters to it.

1. **Introduction**

The dynamics of single-component material surfaces bombarded by energetic ions at off-normal incidence are in general governed by the Cuerno-Barabasi continuum model [1 – 4] which, derived exclusively for ion-sputtered surfaces in an experimental reference frame, is a Kuramoto-Sivashinsky type equation

$$\frac{\partial h}{\partial t} = -v_0 + \gamma \frac{\partial h}{\partial x} + \nu_x \frac{\partial^2 h}{\partial x^2} + \nu_y \frac{\partial^2 h}{\partial y^2} + \frac{\lambda_x}{2}\left(\frac{\partial h}{\partial x}\right)^2 + \frac{\lambda_y}{2}\left(\frac{\partial h}{\partial y}\right)^2 - K\nabla^4 h + \eta \qquad (1)$$

where $h(x,y,t)$ is the height profile at position $(x,y)$ and time $t$, $v_0$ is the erosion velocity for a flat surface, $\nu_x$ and $\nu_y$ are the linear coefficients, $\lambda_x$ and $\lambda_y$ are the non-linear coefficients, $\eta$ is a noise term representative of the stochastic nature of the sputtering process, and $K$ is the surface diffusion coefficient.

This equation reduces to the Bradley-Harper linear theory for small surface slopes and agrees with the Bradley-Harper theory on the formation of ripple topography under certain conditions determined by the parameters of the collision cascades set off by the incident particle as a result of its collision with subsurface atoms upon penetration. The coefficients of equation (1) are functions of the collision cascade parameters and the angle of incidence. When the angle of incidence is zero then the coefficients along the two dimensions are the same and the equation is isotropic.

Additional terms have been incorporated into this equation in a number of cases to account for nanoscale topographies observed in experiments [2, 5 – 8]. Facsko et. al. incorporated an additional linearity to account for short-range hexagonal order [7, 8], while Castro et. al. added another nonlinearity to account for the experimentally observed coarsening of nanodots [2, 5, 9]. According to Castro et. al. [10] such additions are plausible for isotropic sputtering if, in addition to the Barabasi-Stanley recipe for the derivation of the equations of motion [11]; they satisfy the requirement of multi-scale invariance and exploit the relevance of coarsening dynamics.

Other issues arose from an application of the general model or its more robust extensions to many-component systems especially binary compounds [12, 13]. These were before now often attributed to inadequate non-linear parameterization of the strength of the non-linear effects that become increasingly dominant with sputtering time as surface slopes increase. Even in the coupling of specie mobility to the dynamics of the height profile by Munoz-Garcia et. al. [9]; the coupled system was used to evolve the usual, albeit more complex, single dynamical equation of the surface topography evolution with coefficients that are related to experimental sputtering parameters.

However, Bradley and Shipman [14, 15] have now shown that some difficult open questions pertinent to the spontaneous but highly ordered hexagonal nanodots arising from the ion-sputtering of material surfaces made up of binary compounds, in the isotropic case of bombardment at normal incidence, can be answered by using a coupled system of non-linear inhomogeneous partial differential equations. This system models a coupling between the altered composition of a thin surface layer, due to the preferential sputtering of species, and the surface topography which is believed to be the reason behind the hexagonal ordering as well as spontaneous ripple pattern formation on such surfaces.

While this answers a number of open questions on spontaneous nano-pattern formation driven by normal-incidence bombardment, it raises a few more especially as regards off-normal incidence sputtering of the surfaces of binary compounds. Some of these have been addressed in the follow-up paper by Motta, Shipman, and Bradley, on their corresponding oblique-incidence theory [16]. In this paper, we propose an extended model which in addition to the anisotropy in the surface topography [16] also contains anisotropy in the variations of the specie composition as the sputtering proceeds, as necessary consequences of the existing knowledge on continuum models and the Bradley-Shipman theory of surface topography-composition coupling [14, 15]. We then introduce a new scheme with an optimized algorithm which we used to analyze this model in general and discuss its implications.

Before proceeding, note that the coupling treated earlier by Munoz-Garcia et. al. is that between the species transport that arise from the excess energy due to the ion impact and amorphization of a thin layer of the surface and the surface height, whereas that advanced here is the Bradley-Shipman coupling between the varying specie concentration due to the preferential sputtering of one instead of the other and the dynamics of the surface topography. This extends beyond the thin amorphized layer as can be seen from the Bradley-Shipman relation below.

The rest of the paper is organized as follows. In the next section we discuss the aspects of the Bradley-Shipman theory which provide the basis for our work, followed by an introduction of our model. In section III, we introduce the new conformal map scheme and validate it by its application to the Cuerno-Barabasi and Bradley-Shipman models. In section IV, we apply the method to an analysis of our new anisotropic model and present our results. Finally, we conclude in section V and discuss our algorithm for an optimization of the new scheme in the appendix.

**2. The Bradley-Shipman theory and the model of anisotropic specie concentration**

In the paper by Bradley and Shipman [14] the problem of the short-ranged hexagonal order of nanodots was tackled. This problem is a salient feature of the Cuerno-Barabasi non-linearity $\lambda(\nabla h)^2$ or, as the case may be, its higher order extensions and has been the subject of numerous debates and research as to the exact form of the continuum description of a stable long-ranged hexagonal order which can then be studied as a guide and to provide required insight for subsequent experimental verification and validation. Bradley and Shipman tackled this problem with the introduction of another kind of non-linearity that arises from the increasing concentration of surviving specie of atom at the expense of the other which is being preferentially sputtered. Further, this new non-linearity is not just coupled to the dynamics of the surface height since it evolves with time as well, i.e. it has its own dynamics. Hence they modeled the dynamics of these surfaces by coupling the dynamics of the specie concentration to the dynamics of the surface evolution through the following system [14, 15]:

$$\frac{\partial u}{\partial t} = \phi - \nabla^2 u - \nabla^4 u + \lambda(\nabla u)^2 \qquad (2)$$

$$\frac{\partial \phi}{\partial t} = -a\phi + v\phi^2 + \eta\phi^3 + b\nabla^2 u + c\nabla^2 \phi \qquad (3)$$

where the equations have been transformed to dimensionless variables. The specie concentration $c_A$ is expressed in terms of the equilibrium concentration $c_{s,0}$ through the Bradley-Shipman relation

$$c_A = (c_{s,0} - c_b) f\left(\frac{h(t) - z}{d}\right) + c_{s,0} \qquad (4)$$

$c_A$ is the concentration of specie $A$, $c_S$ is the surface concentration, $c_{s,0}$ is the steady state concentration, $d$ is an arbitrary depth in the target, and $z$ is a reference initial depth.

The significance of the coupling is that the height does not only evolve non-uniformly with time as a function of changing surface slopes, due to the curvature dependence of the sputtering yield of a single component, but also evolve non-uniformly with time as a function of specie concentration which varies with time due to the different bonding strengths, such that one species is sputtered preferentially to the other, as well as the slope dependence of the sputtering yield of each specie. That is, if the dependence on specie concentration is ignored, equation (3) vanishes and (2) becomes the dimensionless isotropic and noiseless Cuerno-Barabasi model. Whereas, if the curvature dependence of the surface morphology is ignored (valid at very early times) only the first term on the RHS of (2) remain and the fourth term on the RHS of (3) vanish; even in this case $\Delta c_s = c_s - c_{s,0}$ is so low as for (3) to vanish and the second term on the RHS of (2) replaced by a noisy term.

Based on this argument, existing knowledge of single-component continuum models [2 – 4, 10], and the recent work of Bradley and Shipman [15] we propose the following continuum model for the evolution of surfaces sputtered by off-normal incidence ion bombardment

$$\frac{\partial h}{\partial t} = -v_0 + \Delta c_s + v_x \frac{\partial^2 h}{\partial x^2} + v_y \frac{\partial^2 h}{\partial y^2} + \lambda_x \left(\frac{\partial h}{\partial x}\right)^2 + \lambda_y \left(\frac{\partial h}{\partial y}\right)^2 - \kappa \nabla^4 h \qquad (5)$$

$$\frac{\partial c_s}{\partial t} = -\alpha \Delta c_s + \beta(\Delta c_s)^2 + \gamma(\Delta c_s)^3 + \eta_x \frac{\partial^2 h}{\partial x^2} + \eta_y \frac{\partial^2 h}{\partial y^2} + \mu_x \frac{\partial^2 c_s}{\partial x^2} + \mu_y \frac{\partial^2 c_s}{\partial y^2} \qquad (6)$$

We have assumed isotropic diffusion for simplicity and the coefficients are functions of the experimental sputtering parameters.

In the usual analytic treatment a plane-wave solution of the form

$$H(\mathbf{r}, t) = A(t) \exp[i\mathbf{k} \cdot \mathbf{r}] \qquad (7)$$

is assumed, where

$$h = \frac{H \pm \bar{H}}{2i^{(1\mp 1)/2}} \quad (8)$$

But trying to use this to analyze the continuum equation by complex variable theory leads to unnecessary rigour in the general case of any argument $\theta(r)$ of $H(r,t)$, with the more general amplitude $A(r,t)$, but can be easier when subjected to constraints binding particular experiments. Hence, we provide a numerical solution which can be easily adapted by following the procedure of Ref. [17] which we validate by applying it to the CB and BS models in the next section, after which we then apply it to solve (5) and (6).

### 3. The new method and its solution of the Cuerno-Barabasi and the Bradley-Shipman models.

We adopt a difference scheme [17] for solving the continuum equations like (1), which gives us the discretized version of (1) as

$$h_{i,j,k+1} = h_{i,j,k} + a(h_{i+1,j,k} - h_{i-1,j,k}) + b_x(h_{i+1,j,k} - 2h_{i,j,k} + h_{i-1,j,k}) + b_y(h_{i,j+1,k} - 2h_{i,j,k} + h_{i,j-1,k}) \\ + c_x(h_{i+1,j,k} - h_{i-1,j,k})^2 + c_y(h_{i,j+1,k} - h_{i,j-1,k})^2 - d(h_{i+1,j+1,k} - 2h_{i,j,k} + h_{i-1,j-1,k}) \\ + \eta \quad (9)$$

where $a = \frac{\gamma \Delta t}{2|\Delta x|}, b_x = \frac{v_x \Delta t}{|\Delta x|^2}, b_y = \frac{v_y \Delta t}{|\Delta y|^2}, c_x = \frac{\lambda_x \Delta t}{8|\Delta x|^2}, c_y = \frac{\lambda_y \Delta t}{8|\Delta y|^2}, d = \frac{2K \Delta t}{|\Delta x|^2 |\Delta y|^2}$. A more efficient scheme can be deduced by equating the neglected, higher-order, error terms of (9) as

$$-\frac{|\Delta t|}{2} h''(t_i) \approx -\gamma \frac{|\Delta x|^2}{6} h'''(x_i) - v_x \frac{|\Delta x|^2}{12} h^{(4)}(x_i) - v_y \frac{|\Delta y|^2}{12} h^{(4)}(x_i) - \frac{\lambda_x}{2}\left(\frac{|\Delta x|^2}{6} h'''(x_i)\right)^2 \\ - \frac{\lambda_y}{2}\left(\frac{|\Delta y|^2}{6} h'''(y_i)\right)^2 + 2K \frac{|\Delta x||\Delta y|}{30} h^{(6)}(x_i) \quad (10)$$

and then exploiting the gradient symmetry to express e.g. $\frac{\partial^2 h}{\partial t^2}$ as $\frac{\partial}{\partial t}\left(\frac{\partial h}{\partial t}\right)$, substituting the expression on the RHS of (1) to get partial derivatives w.r.t the spatial coordinates through which the optimal choice of the parameters can be deduced. But this is unnecessary for our purpose here and we simply require that the lattice spacing and time interval be sufficiently small.

However, this difference scheme is unstable for a number of parameter combinations relevant to most experimental conditions. We solve this problem by performing a conformal transformation of the form

$$\mathfrak{S} = f\left(\frac{\partial h}{\partial t}\right)$$

such that the instability, which arises from the dependence of the original scheme on surface gradients which it tends to increase exponentially, is removed. The physical meaning of this new approach is a transformation to an infinitesimal space where all gradients and dynamics are as in the normal space. In our applications of the new method here, it suffices to choose

$$\mathfrak{S} = \varrho \frac{\partial h}{\partial t}; \; \varrho \, \epsilon \, \{k: 0 < k < 1; k \, \epsilon \, \mathbb{R}\}$$

in which case the onset of the instability is delayed beyond the time interval of interest.

Obviously without the noise term the simulation scheme applied to (9), for an (initially) flat surface, leaves the surface flat for all times. This reveals it is an indication that ejected particles may be redeposited back on the erosion spot [$\eta = 0$] or redeposited far enough from the erosion spot such that they bind to the surface elsewhere [$\eta \neq 0$]. For convenience and significance we take $\eta$ to be a uniform deviate of the order of $|\Delta x|$, since it is of the nature of a vertical dimension $|\Delta z|$. Next, the scheme is adapted for memory optimization as shown in the appendix.

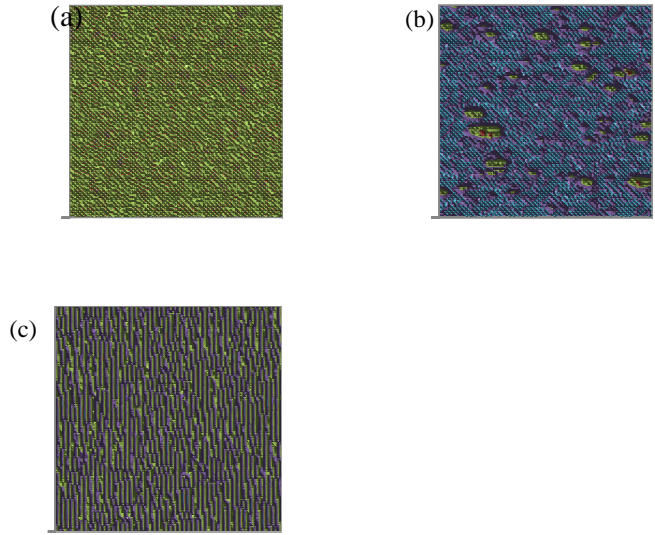

Figure 1: Solutions of the CB model for a variety of values of the CB coefficients for surface nano-structuring by ion bombardment. (a) Ripple structures oriented perpendicular to the ion beam direction; (b) ripples and holes/pits; (c) rotated ripples arising from a swapping of the values of the linear coefficients of (a); (d) non-ripple network-like nano-pattern.

The results obtained by using the new scheme are in good agreement with those obtained from more technical methods. Pertaining to the surface topography the new scheme reveals ripples for certain values of the coefficients [Fig. 1(a)], ripples and holes/pits [Fig. 1 (b)], rotated ripples after a reversal of signs of surface tension coefficients [Fig. 1 (c)], and non-ripple network-like nanopattern [Fig. 1 (d)]. Pertaining to the dynamical evolution of the topography we found morphology involving a transition from roughening at early times [Fig. 2(a)] to ordered ripple structures later [Figs. 2(b) and (c), under continuous ion sputtering. Pertaining to the interplay between the roughening process of sputtering and the smoothing process of surface diffusion the results of the application of the new scheme confirmed that enhanced diffusion delays the emergence of ordered structures; for higher K (lower $|K|$) the ripples are formed much later and even then the ripples are less ordered and less prominent.

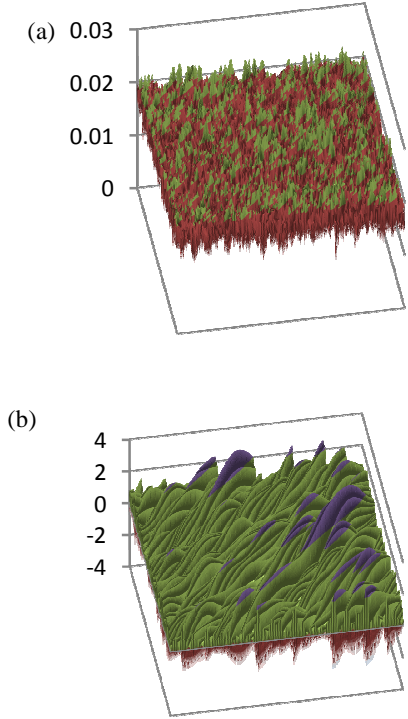

Figure 2: Dynamical evolution of the surface morphology governed by the CB model. (a) t = 10 lattice sweeps; (b) t = 100 lattice sweeps.

We write the discretized version of the Bradley-Shipman model in like manner as the writing of (9):

$$u_{k+1} = u_k + (\phi_{i,j} + \Delta u)\Delta t \quad (11)$$

$$\phi_{k+1} = \phi_k + [P_3(\phi) + f(\nabla u, \nabla \phi)]\Delta t \quad (12)$$

where $k$ is the time index, $i$ and $j$ are space indices,

$$\Delta u = -\left[\left(u_{i+1,j} - 2u_{i,j} + u_{i-1,j}\right)\delta y + \left(u_{i,j+1} - 2u_{i,j} + u_{i,j-1}\right)\delta x + 2\left(u_{i+1,j+1} - 2u_{i,j} + u_{i-1,j-1}\right) \right.$$
$$\left. - \frac{\lambda}{4}\left[\left(u_{i+1,j} - u_{i-1,j}\right)^2 \delta y + \left(u_{i+1,j} - u_{i-1,j}\right)^2 \delta x\right]\right]/\delta x \delta y \quad (13)$$

$$P_3(\phi) = -a\phi_{i,j} + \nu\phi_{i,j}^2 + \eta\phi_{i,j}^3 \quad (14)$$

$$f(\nabla u, \nabla \phi) = \left[b\left[\left(u_{i+1,j} - 2u_{i,j} + u_{i-1,j}\right)\delta y + \left(u_{i,j+1} - 2u_{i,j} + u_{i,j-1}\right)\delta x\right] \right.$$
$$\left. + c\left[\left(\phi_{i+1,j} - 2\phi_{i,j} + \phi_{i-1,j}\right)\delta y + \left(\phi_{i,j+1} - 2\phi_{i,j} + \phi_{i,j-1}\right)\delta x\right]\right]/\delta x \delta y \quad (15)$$

$\delta x = |\Delta x|^2$ and $\delta y = |\Delta y|^2$.

In this case there is no noise term but the presence of the specie concentration term ensures the evolution of the surface even if the initial substrate was flat. We found reasonable agreement with the Bradley-Shipman result and went on to explore parameter combinations outside their sphere with different initial specie distributions. For random specie concentration, no ordering or dots was found.

## 4. Numerical solution of the anisotropic specie composition coupled model

The solution of the discretized version of our anisotropic model is given as follows:

$$h_{k+1} = h_k - \left(v_0 - \Delta c_{s_{i,j}} - \Delta h\right)\Delta t \quad (16)$$

$$c_{s_{k+1}} = c_{s_k} + [P_3(c_s) + f(\nabla h, \nabla c_s)]\Delta t \quad (17)$$

$$\Delta h = \left[v_x(h_{i+1,j} - 2h_{i,j} + h_{i-1,j})\delta y + v_y(h_{i,j+1} - 2h_{i,j} + h_{i,j-1})\delta x - 2\kappa(h_{i+1,j+1} - 2h_{i,j} + h_{i-1,j-1})\right.$$
$$\left. + \frac{1}{4}\left[\lambda_x(h_{i+1,j} - h_{i-1,j})^2 \delta y + \lambda_y(h_{i+1,j} - h_{i-1,j})^2 \delta x\right]\right]/\delta x \delta y \quad (18)$$

$$P_3(c_s) = -\alpha \Delta c_{s_{i,j}} + \beta\left(\Delta c_{s_{i,j}}\right)^2 + \gamma\left(\Delta c_{s_{i,j}}\right)^3 \quad (19)$$

$$f(\nabla h, \nabla c_s) = \left[\eta_x(h_{i+1,j} - 2h_{i,j} + h_{i-1,j})\delta y + \eta_y(h_{i,j+1} - 2h_{i,j} + h_{i,j-1})\delta x\right.$$
$$+ \mu_x\left(\Delta c_{s_{i+1,j}} - 2\Delta c_{s_{i,j}} + \Delta c_{s_{i-1,j}}\right)\delta y$$
$$\left. + \mu_y\left(\Delta c_{s_{i,j+1}} - 2\Delta c_{s_{i,j}} + \Delta c_{s_{i,j-1}}\right)\delta x\right]\frac{1}{\delta x \delta y} \quad (20)$$

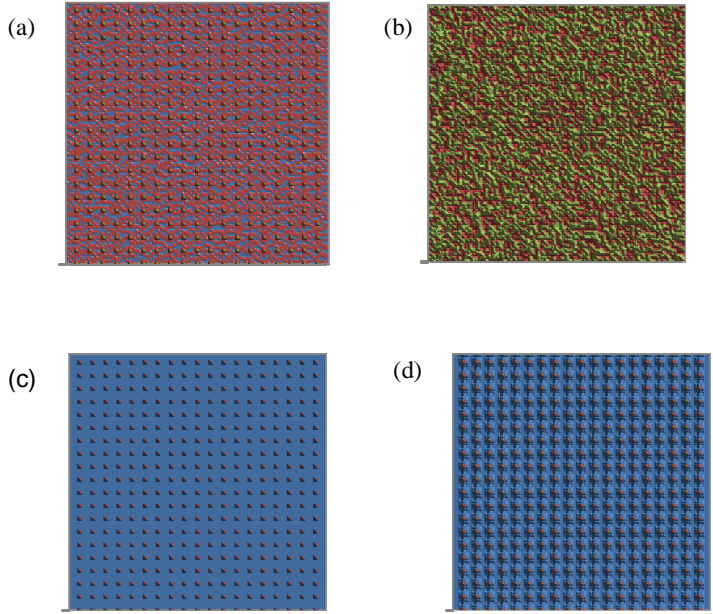

Figure 3: Numerical solutions of the anisotropic model. $a = 0.25, b_x = 3.65, b_y = 0.365, c_x = 1.0, c_y = 5.0, \eta = 10.0, v = 0.0, v_x = -0.04025, v_y = -0.02611, \lambda_x = -0.038, \lambda_y = -0.02811, K = -0.01. t = 1$ lattice sweep for (a) and (c); $t = 200$ lattice sweeps for (b) and (d).

The results are as shown in Fig. 3 for system sizes of 50 x 50. In all results presented an initial distribution of the binary compound structure is in the ratio 3:1 as shown in Figs. 3 (a) and (c) for the initial specie concentration and surface profile, respectively. In the absence of such an initial regular array no dot profile was found. Also, no dot profile [Fig. 3 (b)] was found for a rough initial surface. The coefficients $b_x$ and $b_y$ did not affect the nanodot topography found for the isotropic case of equal $c_x$ and $c_y$, which could be a result of the parameter region in which our values of the coefficients $\nu$ and $\lambda$ falls. In the anisotropic case of $c_x \neq c_y$, the nanodot topography was found to broaden the nanodot topography but with lines connecting neighbouring nanodots. The lines are parallel to the $x$-direction for $\frac{c_x}{c_y} > 1$, and perpendicular to the $x$-direction for $\frac{c_x}{c_y} < 1$ [Fig. 3 (d)].

## 4. Conclusions

We extended, through the inclusion of anisotropy in the variations of the specie concentration, the anisotropic continuum model for surfaces of type $A_xB_{1-x}$ which couples the surface topography to the surface composition. We showed that this model gives the anisotropic Cuerno-Barabasi model of single component surfaces as x approaches unity, and analyzed the general properties of the model. We also introduced a simple numerical method which allows for a fast and accurate simulation of stochastic continuum models. We validated this method by obtaining known results of the solution of two of the most widely applicable continuum models. We finally applied the method to our anisotropic model.

**Acknowledgement**

We thank Professor Mark Bradley for his kind and helpful correspondence.

**Appendix A.**

Adaptation of the numerical scheme for memory optimization.

Assuming the lattice is square we adapt the scheme for memory optimization by defining an index $p = (i-1)L + j$ where $i = 1, 2, \cdots, L$; $j = 1, 2, \cdots, L$; $p = 1, 2, \cdots, L^2$. Using this parameter the $3D$ array is reduced to an $2D$ array where the first dimension relates to the spatial coordinates and the second dimension to time. The $2D$ spatial coordinates indexed in (10) can then be expressed in $1D$ spatial coordinates indexed as $p$ as

$$\begin{aligned}(i,j) = p, (i+1,j) = p+L, (i-1,j) = p-L\\ (i,j+1) = p+1, (i,j-1) = p-1\\ (i+1,j+1) = p+L+1, (i-1,j-1) = p-L-1\end{aligned} \qquad (A1)$$

And (10) becomes

$$\begin{aligned}h_{p,k+1} = h_{p,k} &+ a(h_{p+L,k} - h_{p-L,k}) + b_x(h_{p+L,k} - 2h_{p,k} + h_{p-L,k}) + b_y(h_{p+1,k} - 2h_{p,k} + h_{p-1,k})\\ &+ c_x(h_{p+L,k} - h_{p-L,k})^2 + c_y(h_{p+1,k} - h_{p-1,k})^2 - d(h_{p+L+1,k} - 2h_{p,k} + h_{p-L-1,k})\\ &+ \eta \qquad (A2)\end{aligned}$$

subject to periodic boundary conditions (PBC) (16). The noise term $\eta$ is, by definition, a Gaussian distributed random deviate (11). Noting our above requirement on it, we define it to be

$$\eta = |\Delta x| \cdot r; \quad r \sim N(0,1) \qquad (A3)$$

(10) shows that the interactions do not exceed nearest neighbors, hence, we can implement PBC by letting $i$ and $j$ take integer values from 0 to $L+1$, instead of the actual values 1 to $L$, so that $p = i(L+2) + j + 1$ where

$$\begin{aligned}p(i=0) &= 1, 2, \cdots, L+2\\ p(j=0) &= 1, L+3, 2(L+2)+1, 3(L+2)+1, \cdots, (L+1)(L+2)+1\\ p(i=L+1) &= (L+1)(L+2)+1, (L+1)(L+2)+2, (L+1)(L+2)+3, \cdots, (L+2)^2\\ p(j=L+1) &= (L+2), 2(L+2), 3(L+2), \cdots, (L+2)^2\end{aligned} \qquad (A4)$$

are off-lattice nearest neighbors of the lattice boundaries. This means that the actual $p$ goes from $L+4$ ($i=1, j=1$) to $L(L+3)+1$, exclusive of the $p$ values on the boundaries as defined in $(A4)$. The PBC on the lattice boundaries is then implemented by setting these to the on-lattice nearest neighbors on the other side, before starting a lattice sweep, during which each of the sites is updated only once. Thus the time index $k$ is also an index for a lattice sweep (i.e. the $k$th lattice sweep) and in each time interval the ion beams are rastered over the surface once. With these considerations our final numerical scheme becomes

$$\begin{aligned}h_{p,k+1} = h_{p,k} &+ a(h_{p+L+2,k} - h_{p-L-2,k}) + b_x(h_{p+L+2,k} - 2h_{p,k} + h_{p-L-2,k}) + b_y(h_{p+1,k} - 2h_{p,k} + h_{p-1,k})\\ &+ c_x(h_{p+L+2,k} - h_{p-L-2,k})^2 + c_y(h_{p+1,k} - h_{p-1,k})^2 - d(h_{p+L+3,k} - 2h_{p,k} + h_{p-L-3,k})\\ &+ \eta \qquad (A5)\end{aligned}$$